\documentclass[twocolumn,11pt]{article}
\usepackage{authblk}
\usepackage{hyperref}
\usepackage{amsmath}
\usepackage{graphicx}
\usepackage{dcolumn}
\usepackage{bm}
\usepackage{xcolor}

\usepackage{float}
\usepackage[para]{threeparttable}
\usepackage[a4paper, total={7in, 9in}]{geometry}

\makeatletter
\newcommand{\skipitems}[1]{%
  \addtocounter{\@enumctr}{#1}%
}
\makeatother

\begin{document}
\date{}


    \title{\textbf{High Sensitivity Microwave Spectroscopy in a\\ Cryogenic Buffer Gas Cell}}

\author{Jessica P. Porterfield}
\affil{Harvard Smithsonian Center for Astrophysics, Cambridge, Massachusetts 02138, USA}
\author{Lincoln Satterthwaite}
\affil{Department of Physics, University of California, Santa Barbara, California 93106, USA}
\author{Sandra Eibenberger}
\affil{Fritz-Haber-Institut der Max-Planck-Gesellschaft, Faradayweg 4-6, 14195 Berlin, Germany}
\author{David Patterson}
\affil{Department of Physics, University of California, Santa Barbara, California 93106, USA}
\author{Michael C. McCarthy}
\affil{Harvard Smithsonian Center for Astrophysics, Cambridge, Massachusetts 02138, USA}

\begin{titlepage}
\maketitle
\begin{abstract}
We describe an instrument which can be used to analyze complex chemical mixtures at high resolution and high sensitivity. Molecules are collisionally cooled with helium gas at cryogenic temperatures ($\sim$\,4-7\,K), and subsequently detected using chirped pulse microwave spectroscopy.
Here we demonstrate three significant improvements to the apparatus relative to an earlier version: (1) extension of its operating range by more than a factor of two, from 12\,-\,18\,GHz to 12\,-\,26\,GHz, which allows a much wider range of species to be characterized; (2) improved detection sensitivity owing to use of cryogenically-cooled low-noise amplifiers and protection switches, and (3) a versatile method of sample input that enables analysis of solids, liquids, gases, and solutions, without the need for chemical separation (as demonstrated with a 12\,-\,16\,GHz spectrum of lemon oil).
This instrument can record broadband microwave spectra at comparable sensitivity to high $Q$ cavity spectrometers which use pulsed supersonic jets, but up to 3000 times faster with a modest increase in sample consumption rate.
\end{abstract}
\end{titlepage}

\section{Introduction}
\vspace{-3mm}
Microwave spectroscopy is a powerful tool for chemical analysis, and provides our most accurate measurements of molecular structure.  
It is often performed at low temperatures since the rotational partition function scales as T$^{3/2}$, meaning that for relatively large molecules (e.g., $>$10 atoms), spectral congestion at higher temperatures becomes unmanageable due to occupation of many possible ro-vibrational states.
By far the most widely used source of cold ($\sim$1\,-\,5\,K) molecules is the pulsed supersonic jet \cite{mccarthy2000microwave}.
All spectrometers based on pulsed jets operate at a low overall duty cycle, typically of order 0.1\%. This constraint arises from the large amount of carrier gas ($\sim$99\%) required for efficient rotational cooling, which overwhelms even large vacuum pumps if the pulsed jet is operated at a repetition rate of more than 10\,Hz.  

Microwave spectrometers based on pulsed supersonic jets can be operated as either a \emph{cavity-enhanced instrument}  or a \emph{broadband instrument}. The cavity-enhanced, Fabry-Perot style instrument achieves high sensitivity, but has an intrinsically narrow instantaneous frequency bandwidth (IFBW) of about 1\,MHz \cite{balle1981fabry,mccarthy2000microwave}. The cavity must be re-tuned many times to acquire a broadband spectrum, a process which becomes highly inefficient for wide spectral scans. In contrast, the broadband configuration commonly referred to as a chirped pulse Fourier transform microwave (CP-FTMW) instrument \cite{brown2008broadband,schmitz2012multi,perez2013broadband,crabtree2016microwave}, achieves  wide frequency coverage at the expense of sensitivity. In CP-FTMW spectroscopy it is now straightforward to generate and amplify a linear frequency chirp, owing to rapid improvements in high-speed digital electronics, and broadcast it between two microwave horns where the gas sample resides; the resulting molecular free induction decay (FID) is then directly digitized by a fast digital oscilloscope.
 
We present here a broadband instrument that is capable of achieving signal to noise ratios (SNRs) comparable to that of a cavity instrument, but 3000 times faster per unit time.
Molecules introduced to the experiment are cooled continuously via collisions with cryogenic (4\,-\,7\,K) helium \cite{willey1988very}, a process called \emph{buffer gas cooling}, which enables the data acquisition rate to approach unity.
Our instrument achieves high sensitivity via three advantages over conventional pulsed jet instruments. First, with a continuous source of cold molecules our repetition rate is limited only by the length of the molecular signal, which is set by the molecule-helium collision rate to be on the order of 10 $\mu$s. Typical repetition rates are 25\,-\,30\,kHz, with demonstrated rates as high as 50\,kHz at higher buffer gas densities. Second, sample input rate is no longer limited by a pulsed valve, and thus much higher sample input rates can be used to obtain higher SNRs per unit time. Third, the first-stage amplifier and protection switches are cryogenically cooled, allowing for an instrument noise temperature close to the fundamental limit (T$_{\textrm{sys}}$ $\sim$\,30\,-\,35\,K), as opposed to pulsed jet experiments which have a typical noise temperature of 300\,K or greater. 

An instrument which couples CP-FTMW spectroscopy with cryogenic buffer gas cooling was the subject of a previous publication by one of the authors \cite{patterson2012cooling}.
That instrument was used with considerable success to demonstrate enantiomer specific rotational detection  \cite{patterson2013enantiomer,patterson2015slow} and enantiomer specific state transfer of chiral molecules  \cite{eibenberger2017enantiomer}.
Other methods of detection, such as infrared frequency comb \cite{spaun2016continuous}, CP-millimeter wave spectroscopy \cite{zhou2015direct}, and laser induced fluorescence \cite{piskorski2014cooling,patterson2010cooling} have also recently been coupled to buffer gas cooling cells. 
The instrument described here was recently used to analyze product ratios in a gas phase chemical reaction \cite{porterfield2018ozonolysis}, as well as to record rotational spectra of gases evaporated directly from solid precursors \cite{joost2018sulfur}. 

In this article we describe this next-generation instrument in detail. Substantial improvements in bandwidth, improved detection sensitivity, and a highly versatile method for sample introduction to study solid, liquid, and gaseous samples are described. Improvements in detection sensitivity and SNR per unit time are illustrated with sample spectra of the complex mixture lemon oil and detection of $^{18}$O$^{13}$C$^{34}$S in natural abundance (0.00009\%).
Comparisons of detection sensitivity and IFBW for cavity, traditional broadband, and buffer gas CP-FTMW spectrometers are also provided.  

\vspace{-3mm}\section{The Buffer Gas Cell}
\begin{figure}[t!]
\includegraphics[width=0.48\textwidth]{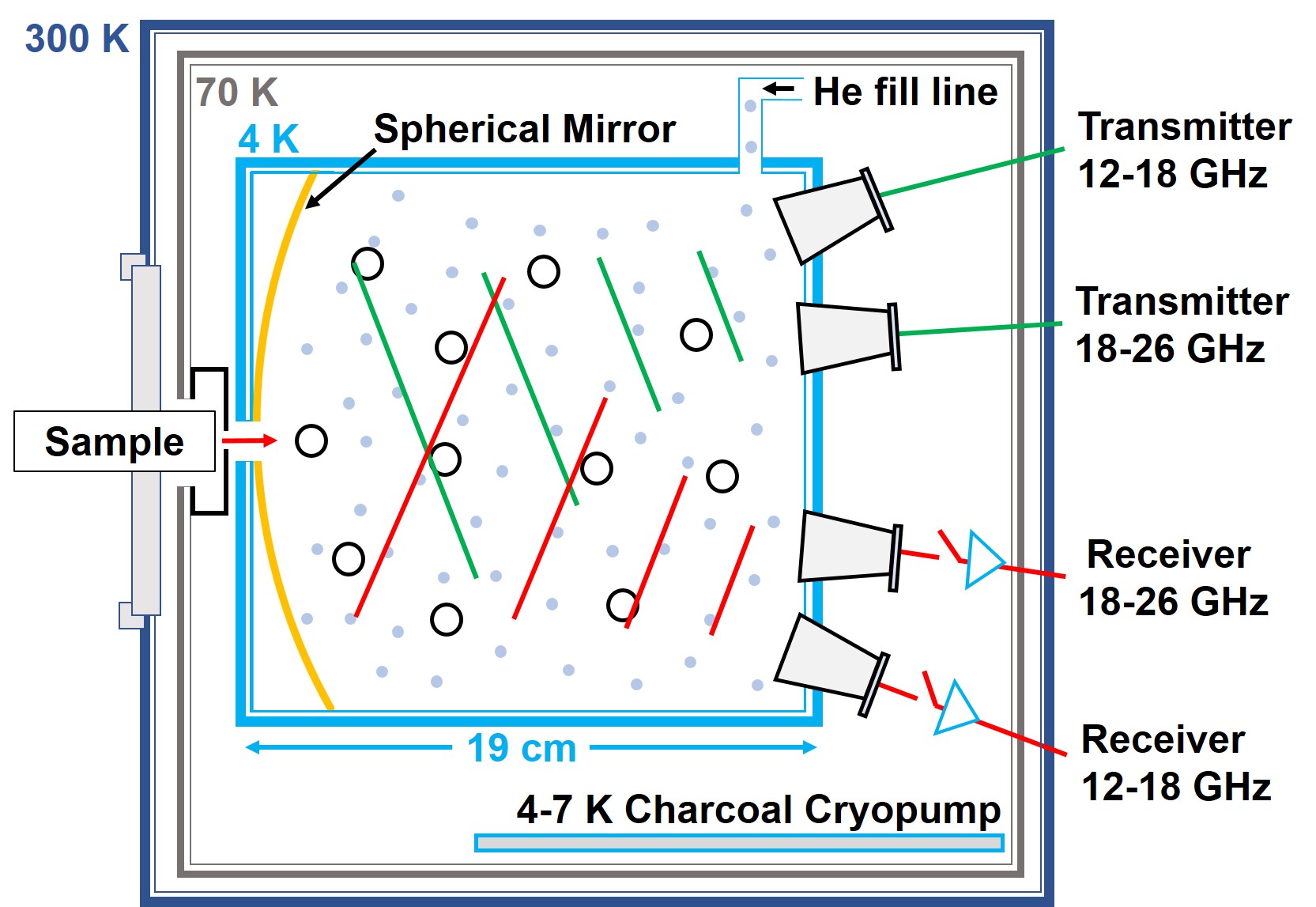}
\caption{A schematic of the chamber containing the cryogenic buffer gas cell. The charcoal cryopump, microwave antennas, receiver protection switch, and buffer gas cell are all thermally anchored to the second (4\,K) stage of the closed cycle cryocooler. The aluminum radiation shield is held at 70\,K by the first stage of the cryocooler. The dimensions of the 4\,K buffer gas cell are 19 x 19 x 19\,cm. For clarity, antennas have been placed in this diagram and in Figure \ref{circuit} on what appears to be a straight line on one side of the cell. Their true geometry, however, is actually a diamond configuration on the back wall of the cell, with the 12-18\,GHz and 18-26\,GHz circuits on the left/right and top/bottom, respectively.}\label{BGC}
\end{figure}

\vspace{-3mm}A schematic of the present apparatus is presented in Figure \ref{BGC}. The cell is cooled using a two-stage closed cycle cryocooler (Cryomech CP980 compressor and PT410 cold head), where the first stage  reaches approximately 70\,K and the second 4-7\,K. 
Inside the apparatus, a 0.125'' thick aluminum radiation shield is thermally anchored to the 70\,K stage by a copper breadboard (Super-Conductive 101 copper). 
Aluminized Mylar superinsulation provides radiation shielding between the room temperature vacuum chamber walls and the 70\,K aluminum radiation shield. 
Inside the radiation shield lies the buffer gas cell (19 x 19 x 19\,cm 6061 aluminum) and cryogenic electronics (amplifiers and switches), all of which are thermally anchored to the second stage of the cryocooler.
Helium gas enters from the top of the cell via a pre-cooled line (4\,K) at a typical flow rate of 7\,standard cubic centimeters per minute (sccm), which results in a density of approximately 1 x 10$^{14}$ He atoms/cm$^{3}$ in the cell. 
An aperture (19\,mm in diameter) allows for molecular input through a hole in the spherical mirror on one face of the cell. 
Molecules collisionally thermalize in nearly all internal degrees of freedom with the 4\,K helium gas within roughly 100 collisions, the exact number depending somewhat on the size of the molecule  \cite{patterson2012cooling, porterfield2018ozonolysis}.
Given the typical helium cell density,
collisions of the molecules with a helium atom occur once every 10-20\,$\mu$s, meaning that they reach equilibrium with the cold helium gas within about 1\,ms. Because the diffusion time to the cell walls is 15-20\,ms, there is a relatively long observation time prior to freezing out on the cell walls  \cite{patterson2012cooling, porterfield2018ozonolysis}.

\begin{figure}[t!]
\centering
\includegraphics[width=0.5\textwidth]{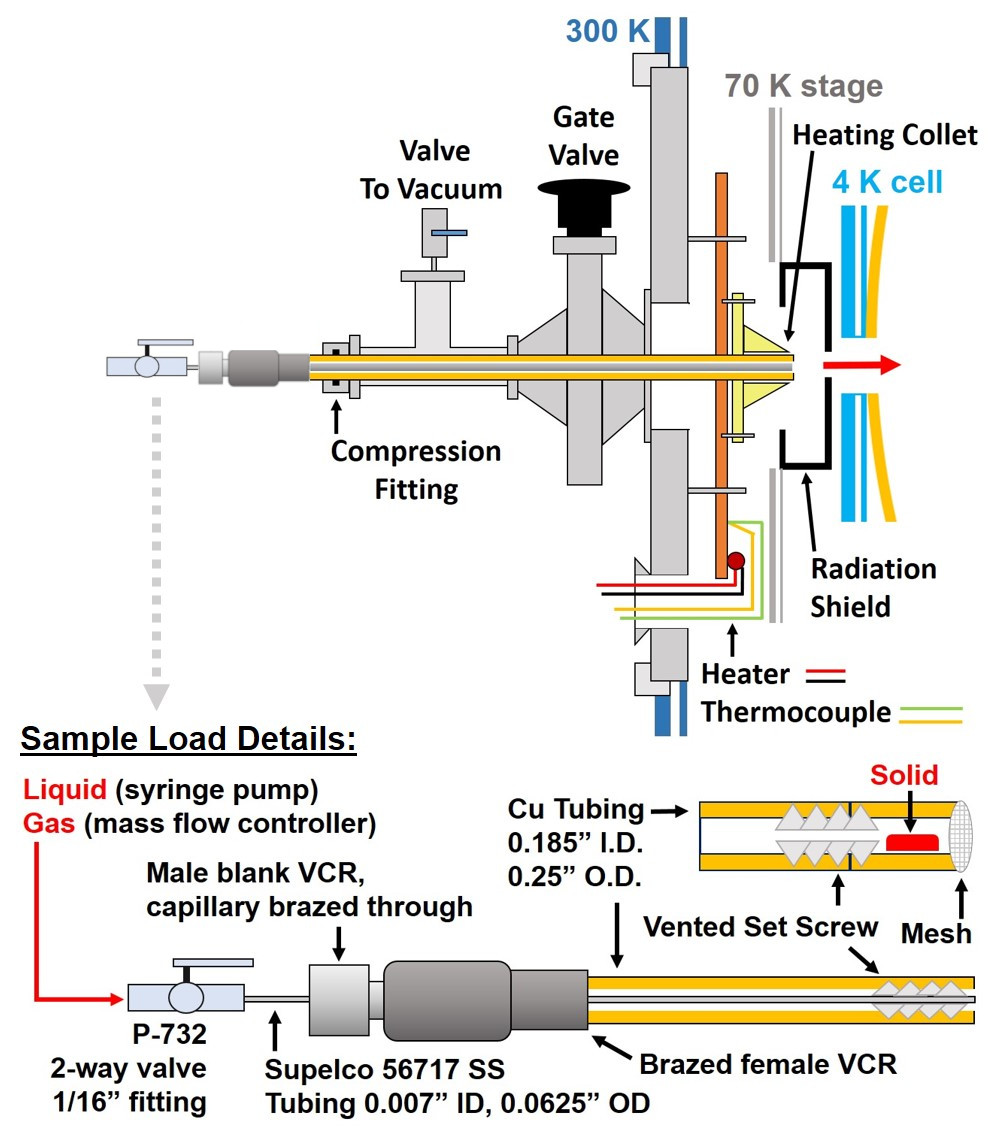}
\caption{A cross-sectional schematic of the sample introduction source, with slightly different configurations for solids, liquids, and gases to be introduced to the cryogenic 4 K buffer gas cell.}\label{samples}
\end{figure}

Although a sample must be in the gas phase when it enters the buffer gas cell, it can be introduced through the cell aperture in multiple ways depending on its initial phase, as illustrated in Figure \ref{samples}.
A  copper 0.250'' diameter sample injection tube passes through a 1 cm diameter aperture in the radiation shield. The end of the sample injection tube is approximately centered and coplanar with this aperture.
This arrangement allows the warm, gas-phase sample to be situated relatively close to the cell ($<$\,2 cm), while shielding the 4\,K stage from excess black body radiation.
If no heating were provided to the sample source, cold helium escaping the buffer gas cell would cool the sample injection tube tip and eventually freeze the flow of sample.

The sample introduction source is a collection of heating and centering pieces attached to a vacuum flange (ISO200) mounted on the outside wall of the chamber (Figure \ref{samples}).
A copper breadboard is held by stainless steel thermal standoffs 12 mm from this flange, and holds a circular brass collet.
The collet heats and centers the 1/4" copper sample injection tube, which is aimed directly at the buffer gas cell entrance.
This copper tube serves as a heating jacket for a stainless steel capillary (Supelco 56717, 0.007" I.D., 1/16" O.D.), which is used to deliver the sample to the cell.
Heat is supplied directly to the copper breadboard and brass collet via two 50 W resistive cartridge heaters; these in turn heat the copper tubing and stainless steel capillary.
Thermal contact is maintained between the copper tubing and capillary utilizing a 1/16" inner diameter vented set screw.
Radiative coupling is sufficient to maintain T$_{\textrm{capillary}}$ $\approx$ T$_{\textrm{tube}}$.
Type-K thermocouple and heater wires are run from the copper breadboard to a KF-25 electrical feedthrough on the ISO200 flange.
Typical temperatures range from 40-100\,$^{\circ}$C at the end of the sample injection tube (180\,$^{\circ}$C has been achieved). The sample injection tube assembly can be replaced without breaking vacuum and without warming up the cryocooler, as described below.

Gaseous samples are typically introduced through the heated stainless steel capillary at 1-10 sccm with a mass flow controller (e.g. MKS series GM50A, 10 sccm N$_2$). Liquid samples are injected with a syringe pump (Aladdin AL-4000 pump) and standard high pressure liquid chromatography fittings, with flow rates of 1-10\,$\mu$L/min. 
The liquid is heated as it is pushed down the capillary, and vaporizes before being introduced into the cell.
To interface the syringe (Hamilton Gastight \#1001, 1 mL) or flow controllers to the input manifold, the capillary and copper are connected with standard VCR fittings. 
In this case, the 1/16" stainless steel capillary is brazed through a male VCR 1/4" blank, and the copper 1/4" tubing is brazed to a female VCR gland.

Solid sample (typically powder) is introduced with a sample holder tip, which is an extension of additional 1/4" tubing (0.180" I.D.).
The sample holder tip is typically 15 mm in length, and is capped with mesh at the exit to prevent the solid from escaping. 
A stainless steel vented set screw is used to attach the sample holder tip to the rest of the heated 1/4" copper tubing.
Use of a solid sample was most recently demonstrated with 7-sulfinylamino-7-azadibenzonorbornadiene, which upon heating was shown to release sulfur monoxide (SO) \cite{joost2018sulfur}. 
Solids with vapor pressures as low as 15 mTorr have been successfully coupled to the buffer gas cell using this introduction technique \cite{asselin2017characterising}.

It is highly desirable to be able to switch the sample input while the system is cold, either for cleaning, recharging solid samples, or changing the injection method. 
To maintain vacuum during this change over, an airlock system is used.
The airlock consists of a gate valve connected to a KF-25 tee, with the 90$^{\circ}$ branch connected to a rotary vane pump, and the straight branch to a  1/4" compression fitting.
To change the sample, the sample tube is withdrawn through the compression fitting from the collet past the gate valve, but not so far as the compression fitting in order to maintain vacuum.
The gate valve is then closed, and the injection tube is withdrawn completely.
For reinsertion, the tube is placed back into the compression fitting, the airlock is pumped out, the gate valve is re-opened, and the sample tube is reinserted into the collet.

To allow for continuous flow of cold helium and sample into the cell, an efficient means of pumping has been employed using a `sorb' or charcoal cryopump.
The sorb is composed of three stacked copper plates (20 x 20\,cm, 0.3\,cm thick middle plate, 0.15\,cm thick outer plates, Super-Conductive 101 copper), spaced by 1/2", and thermally anchored to the second stage of the cryocooler. 
Each of these plates is coated in activated coconut charcoal (8-30 mesh, Spectrum Chemical C1221), which makes thermal contact with the copper plates via epoxy.
This epoxy has a coefficient of thermal expansion to match that of copper (Stycast 2850 epoxy, Henkel Loctite catalyse 24LV curing epoxy hardener). 
The grain size and type (activated coconut) of charcoal are of critical importance in this application, with very fine grain charcoal becoming coated and lost in the epoxy, and very large grain having too little surface area for efficient pumping. 
Pumping speeds achieved by the sorb are up to 6.7 L/s$\cdot$cm$^2$ for helium at 6 K \cite{tobin1987evaluation}. 
To efficiently cryopump helium, the charcoal must remain below a temperature of 8.5 K. 
If this temperature is exceeded, the weak binding of helium gas to the charcoal sorb is broken, and the helium is desorbed. 
This desorption leads to a `crash,' or a thermal coupling of the 300 K vacuum chamber and the cryogenic stage, which immediately overloads the cryocooler and forces a thermal cycling of the system to regenerate the sorb.
When the sorb is brand new, typical run times at 7 sccm helium input and 2 sccm sample input are up to 20 hours.
Over time the sorb becomes saturated more quickly, likely owing to contaminants and water irreversibly adhering to the charcoal, and it therefore must eventually be replaced.
We have substantial evidence that some compounds, most notably fluorobenzene \cite{patterson2012cooling}, dramatically shorten the lifetime of the sorb. 

The temperatures throughout the cryocooler are monitored by silicon diode thermometers (Lakeshore DT-670A1-SD) which are connected to a Lakeshore 218 temperature monitor. 
There are multiple critical locations, namely (1) the buffer gas cell, near the sample input, (2) the charcoal sorb, (3) near the first stage cryogenic amplifiers. 
The temperature of the sorb provides a sensitive indication on the overall health of the system, with subtle changes in reading due to helium pressure and overall gas input into the cell. 
Because the first-stage amplifiers determine the noise floor of our measurements, their temperature is also important. 
Up to six additional thermometers are positioned at other points in the cell for various diagnostics, e.g., on the second stage of the cryocooler, outer radiation shield, or buffer gas cell. 
\vspace{-3mm}\section{Microwave Electronics}
\vspace{-3mm}Microwave spectroscopy uses the intrinsically narrow, electric-dipole allowed rotational transitions that are present in all polar molecules to record highly specific ``fingerprints'' of molecular mixtures. As in nuclear magnetic resonance, molecules are polarized by a short applied field, and the resulting induced coherent oscillation, or free induction decay (FID), of the molecular ensemble is recorded. In NMR the applied and induced fields are magnetic; in microwave spectroscopy they are electric.  A Fourier transform of the FID produces a spectrum. Here, as in most microwave spectroscopy experiments, many individual FIDs are summed in the time-domain prior to taking the Fourier transform. 

Our broadband spectra are assembled from a series of relatively narrow segments, each with a bandwidth of 200 MHz. Each segment is generated by upconverting a chirped waveform produced by mixing a local oscillator  with an arbitrary waveform generator. 
In this experiment, a synthesizer (HP8673D) was used as the local oscillator (L.O.); it is tunable over the experimentally required 12-26\,GHz range, and has very high spectral purity.
The chirp produced from the arbitrary waveform generator (AWG) typically spans frequencies from 25\,-\,125\,MHz. 
Upon mixing with the L.O., the resulting upconverted chirp is then amplified by a gated solid state amplifier. The final chirp has a power entering the cryocooler of $\sim$32\,dBm, and a duration of $\sim$5\,$\mu$s.

\begin{figure}[ht]
\centering
\includegraphics[width=0.45\textwidth]{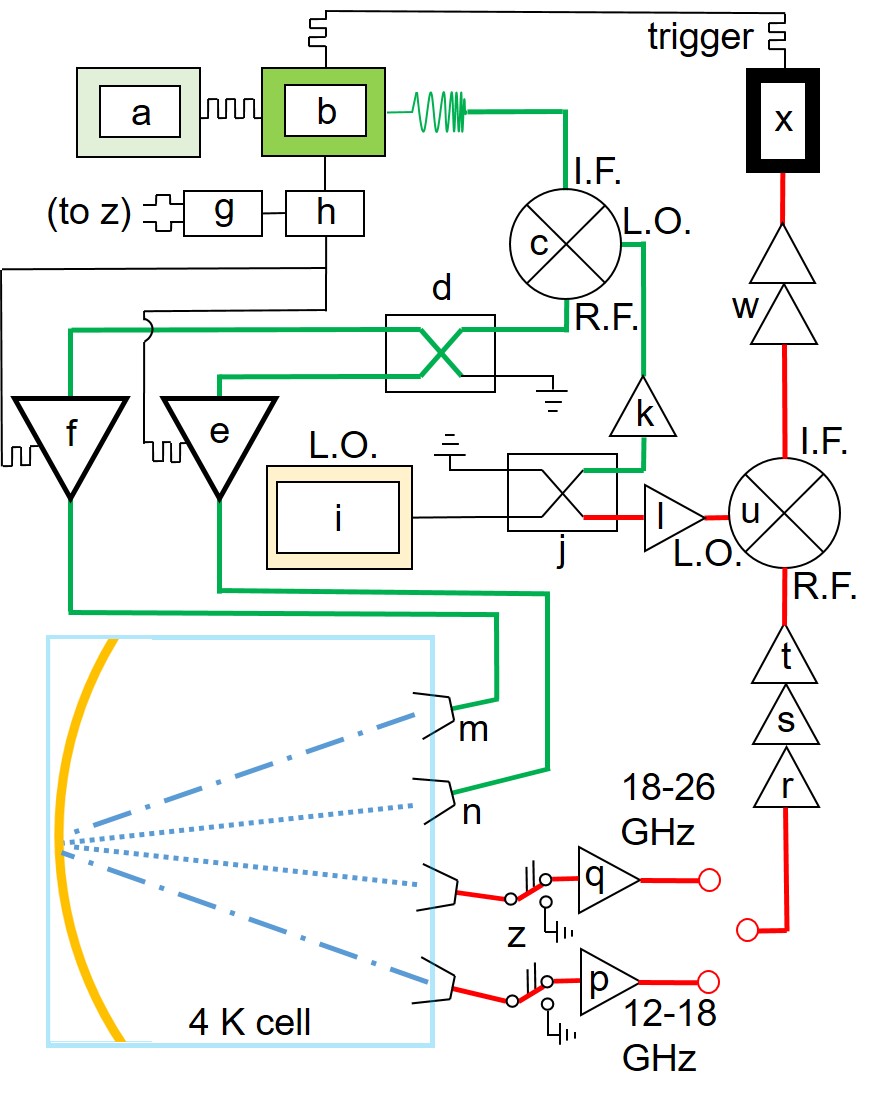}
\caption{The microwave electronic circuit diagram. Letters indicate select parts of the circuit that are discussed within the Microwave Electronics section. The transmission and emission circuits are represented by green and red lines on the diagram, respectively. Enhanced sensitivity is attributed to the low thermal noise of the cryogenic amplifiers (q,p) and protection switches (z), in addition to the overall allowable repetition rate (up to 50\,kHz). The 12\,-\,18\,GHz circuit (f,m) is separate from the 18\,-\,26\,GHz circuit (e,n).}\label{circuit}
\end{figure}    

\begin{figure}[ht!]
\centering
\includegraphics[width=0.48\textwidth]{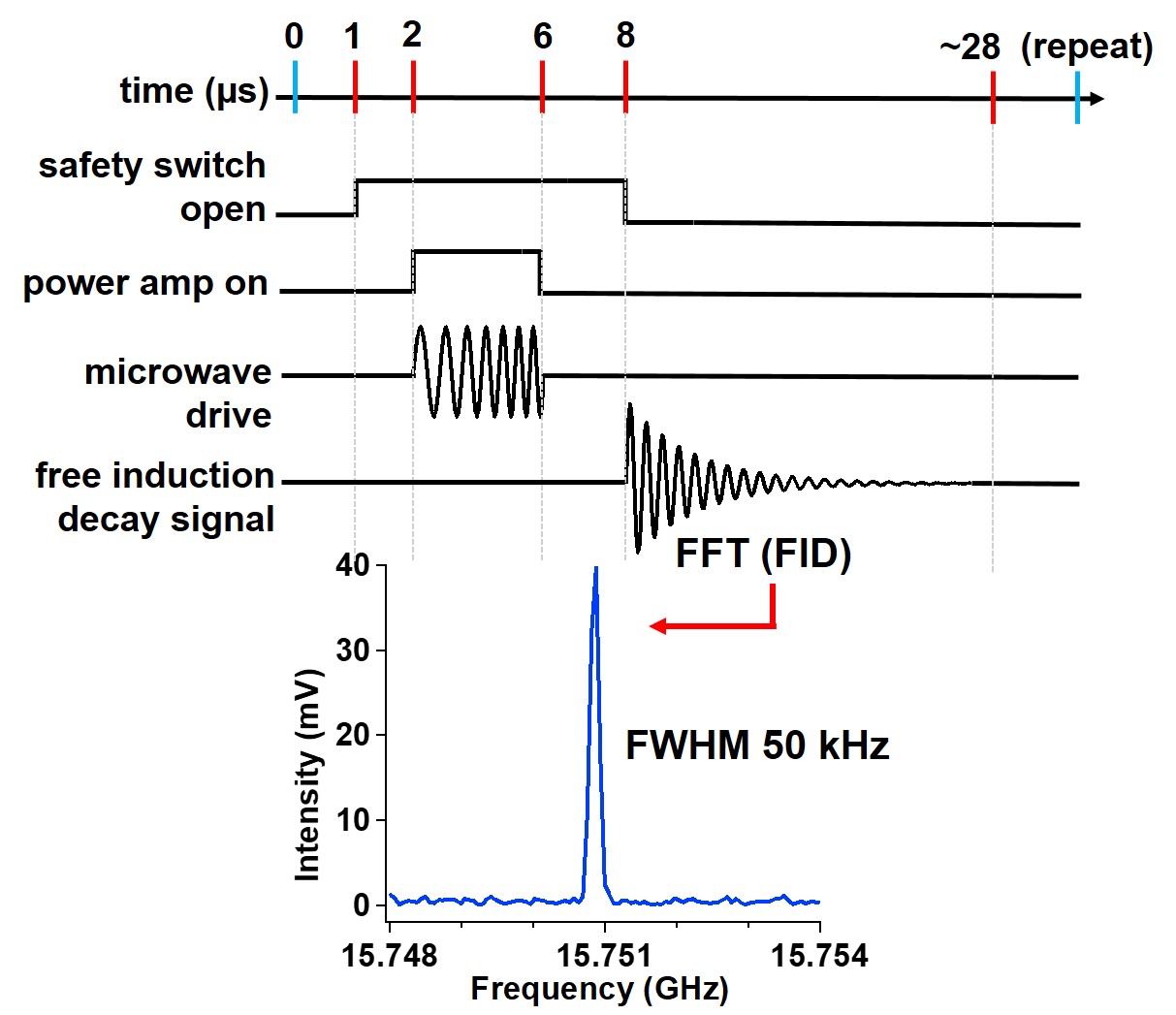}
\caption{A typical timing diagram including relative positions in $\mu$s of the safety switch open time, power amplified chirp transmission, and free induction decay molecular signal, which is Fourier-transformed into a frequency spectrum. The cycle is repeated at approximately 30\,-\,50\,kHz, with FID collection lasting between 5 and 20\,$\mu$s.}\label{timing} 
\end{figure}
    
A schematic of the microwave circuit is presented in Figure~\ref{circuit}, with lowercase letters indicating components of the circuit.
A list of the components of our microwave circuit, as well as their important attributes, can be found in the Supplementary Information following the labels presented in Fig.~\ref{circuit}.
A timing diagram is presented in Figure \ref{timing}. The 12\,-\,18\,GHz and 18\,-\,26\,GHz circuits are operated separately due to the need for distinct circuit components for different frequency regimes (namely 12-18\,GHz f,m,p; 18-26\,GHz e,n,q). 
The overall timing is controlled by a homemade programmable TTL pulse generator (a), which triggers the generation of a chirp pulse from the AWG (b). 
The AWG then triggers the U1084A data acquisition card (x), and simultaneously triggers the delay generator (h). 
The delay generator then triggers the power amplifier (either 12\,-\,18\,GHz or 18\,-\,26\,GHz, depending on mode of operation; e, f) and protection switches (g,z). 
The output of the L.O. (i) is split (j) and then amplified for the transmitter (k) and receiver (l) circuits. 

In the transmitter circuit, the chirp from the AWG is mixed (c) with the L.O., and the signal is split (d) into the 12\,-\,18\,GHz (f,m,p) and 18\,-\,26\,GHz (e,n,q) circuits. The pulse is broadcast from a horn antenna (m,n) and is directed through molecules in the 4\,K cell towards a spherical mirror, which reflects and focuses the microwaves back onto a second horn antenna.
Following excitation, the molecules emit a coherent, oscillating electric field, which propagates along the same path as the excitation pulse. This  signal passes a protection switch (z) and is detected by the cryogenic low noise amplifiers (LNAs; p, q). 
Depending on whether the instrument is in 12\,-\,18\,GHz or 18\,-\,26\,GHz mode, the output of the respective LNA is fed directly into a series of room temperature amplifiers (r,s,t) and then mixed down (u) with the amplified output of the L.O. (l). After the mix down step, the signal is amplified (w), and fed to the U1084A data acquisition card (x).

Spectral resolution and overall repetition rate are both a function of helium gas density and chirp conditions.
The FID lifetime, and hence our resolution, is set by collisional broadening within the cell. 
At higher buffer gas densities the instrument has been run up to 50 kHz, however more standard operation is in the 25\,-\,35\,kHz regime. 
Higher resolution can be achieved at the expense of signal amplitude by using a lower helium flow, which results in less pressure broadening. The strength of the chirp can be controlled via the chirp duration and bandwidth.
We observe modestly shorter FIDs for large molecules, which exhibit larger helium-molecule inelastic cross sections, and hence larger collisional broadening.

A major challenge in developing this instrument was identifying a protection switch which operated at cryogenic temperatures. 
In contrast to LNAs, which are widely used by the radio astronomy community, we know of no wide use case for cryogenic switches.  
However, we found that the HMC547ALC3 family seem to work reliably at cryogenic temperatures, despite being rated for operation only at -40$^{\circ}$\,C and above.  We believe that this switch works well because it contains only GaAs components; in particular, in contrast to most commercially available switches (e.g., NARDA S123D, which does not work cryogenically) it contains no silicon TTL drivers. We have not carefully characterized the noise temperature of this switch at low temperatures, although we have evidence that it is 25\,K or lower.

The instrument can be operated in one of two modes. In \emph{single segment mode}, the local oscillator is fixed, and a single spectral segment (200\,MHz bandwidth) is recorded (Fig.~\ref{OCS}). This mode is appropriate when observing a single transition, and can be most directly compared to a cavity-enhanced pulsed jet spectrometer. In \emph{survey mode}, a broadband spectrum is stitched together from many single segments (Fig.~\ref{lemons}). This mode can be most directly compared to a CP-FTMW pulsed jet spectrometer.
Although the IFBW of the instrument could in principle be increased further, careful consideration must be given to the overall heat load and cooling capacity of the cryocooler.
The cooling power of the PT410 is about 1.0\,W at 4.2\,K, and the circuit described here has time-averaged output power delivered to the cell of about 100\,mW. 
To get the same sensitivity at double the bandwidth, for example, would require twice the amount of microwave power, thereby adding an additional 100\,mW to the heat load on the cell.
Caution is therefore appropriate when significantly extending the bandwidth of the instrument presented here.


No attempt at single sideband (SSB) upconversion or downconversion is made, so an observed signal at intermediate frequency $f$ could be a molecular signal at $f_{LO} + f$ or $f_{LO} - f$.  Although in principle SSB downconversion would resolve this ambiguity, our high dynamic range means that an ambiguity between a strong signal in the lower sideband and a weak signal in the upper sideband would remain.  We resolve this ambiguity  when operating in survey mode by recording enough spectra at distinct local oscillator frequencies such that each frequency in our broad spectrum is covered multiple ($\sim10$) times. A true peak will appear in all overlapping spectra, while a peak from the wrong sideband will appear in just one.  After testing several algorithms for combining the overlapping spectra, we found that a simple geometric mean provided robust performance and low noise  \cite{Peebleschirp}. A typical broadband spectrum is composed of a few hundred such overlapping segments, with local oscillator frequencies spaced by $\sim$\,50\,MHz. Each segment is recorded for roughly 20 seconds.
\vspace{-3mm}\section{Discussion}
\begin{figure}[t!]
\centering
\includegraphics[width=0.5\textwidth]{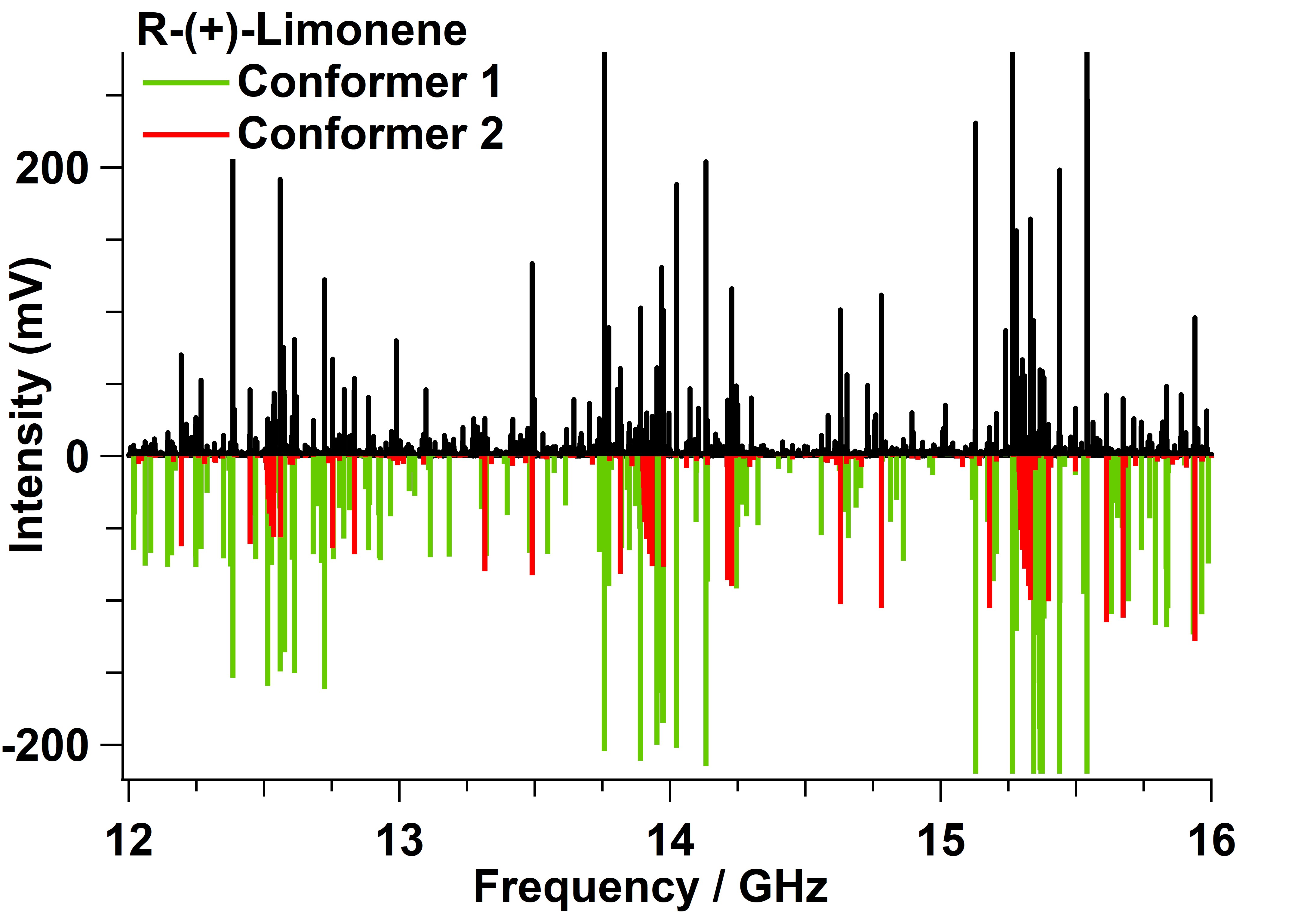}
\caption{A sample spectrum of lemon oil from 12-16 GHz recorded with a helium flow of 5 sccm and a sample input rate of 3.5 $\mu$L/min. The spectrum is assembled from 75 individual spectral segments, with 80 MHz local oscillator steps between them. Each segment was taken with 200 MHz bandwidth, a 4$\mu$s long chirp, and 15 seconds of integration. Clear evidence of the two most stable conformers of limonene are apparent. Lines of $\beta$-pinene as well as trace contaminants such as ethanol (parts per thousand scale) were also found in this frequency range.}\label{lemons}
\end{figure}

\begin{figure}
\centering
\includegraphics[width=0.45\textwidth]{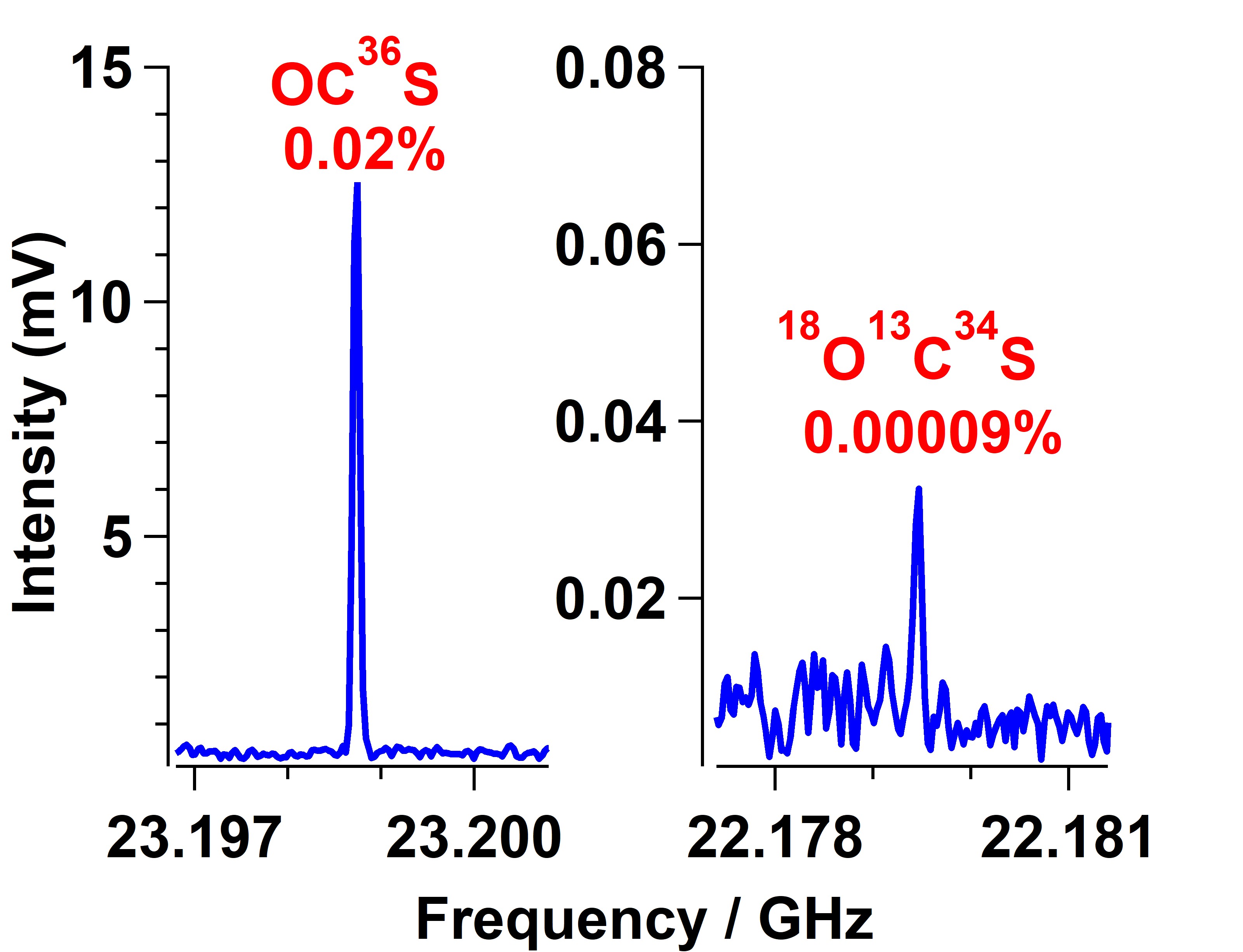}
\caption{Two sample spectra of OCS is natural isotopic abundance demonstrating the sensitivity achieved with the present instrument. The OC$^{36}$S species is composed of 500,000 averages and the $^{18}$O$^{13}$C$^{34}$S 70 million averages, with 20 and 3,000 seconds of integration, respectively.}
\label{OCS}
\end{figure}

\vspace{-3mm}We present two experiments that illustrate the advantages of this instrument: a 12\,-\,16\,GHz spectrum of lemon oil (Figure \ref{lemons}), and detection of rare isotopic species of carbonyl sulfide (e.g., $^{18}$O$^{13}$C$^{34}$S; Figure \ref{OCS}) in natural isotopic abundance. 
The spectrum of lemon oil demonstrates our ability to characterize a complex liquid solution and determine its content without chemical separation.
The 12\,-\,16\,GHz spectrum was acquired with 35\,min of integration.
Lemon oil is composed of over 25 compounds \cite{veriotti2001high}, and we are able to assign the most prevalent component limonene (C$_{10}$H$_{16}$) with high signal to noise, as well as $\beta$-pinene and trace contaminants such as ethanol on the parts per thousand scale in a straightforward and confident manner. 
No attempt was made to determine relative quantities of species from the spectrum in Fig.~\ref{lemons}, but given the large number of transitions observed in this region, this can almost certainly be done to within a 20\% margin of error \cite{porterfield2018ozonolysis} based on the relative intensities. 

Conformational cooling can vary when comparing cryogenic collisional cooling to pulsed supersonic jets \cite{drayna2016direct}.
In the case of lemon oil, however, two of the three calculated conformers of R-(+)-limonene were observed in our experiment (displayed in red and green on the negative y-axis in Fig. \ref{lemons}), as was observed in previous studies of limonene in a supersonic jet observed by FTMW spectroscopy \cite{moreno2013conformational} and photoelectron circular dichroism using single photon ionization \cite{rafiee2018intense}.
All three conformers of limonene are expected to have comparable dipole moments ($\sim$0.3 D), however the highest energy conformer is yet to be observed \cite{moreno2013conformational}.

A natural figure of merit when comparing spectrometers of different designs is the \emph{spectral acquisition velocity} $S$ (MHz/min), defined as the product of the single line sensitivity and the IFBW normalized with respect to time.
To precisely quantify $S$ for the present instrument in relation to both cavity and CP-FTMW spectrometers using pulsed jets, rotational lines of rare isotopic species of OCS \cite{kubo2003submillimeter} were detected. Table~\ref{OCStable} summarizes the results of these measurements, and Fig.~\ref{OCS} displays two of the OCS lines that were observed under identical conditions (0.25\,sccm OCS, 7\,sccm He) in the cyrogenic buffer gas cell.

With respect to the two pulsed jet experiments, the  single line detection sensitivity of the cavity variant is roughly 100 times greater than that of the CP version. This sensitivity difference is  consistent with the expected $Q^{1/2}$ dependence \cite{park2016perspective}, where $Q$ is the quality factor ($Q= \nu/\Delta \nu$, with $Q \approx  2 \times 10^4$ vs.~$Q=1$ for the cavity and CP, respectively).  $S$, however, is  considerably larger for the CP variant, owing to its much larger IFBW, which illustrates the significant advantage of this type of spectroscopy if wide spectral coverage is required.

\begin{table*}
\centering
\caption{Comparison of two OCS isotopic species detected in natural abundance with the buffer gas cell (bgc), and traditional cavity and broadband spectrometers employing a pulsed jet nozzle source.}\label{OCStable}
\begin{threeparttable}
\begin{tabular}
{l@{\hspace{1em}}c@{\hspace{1em}}c@{\hspace{1em}}c@{\hspace{1em}}r@{\hspace{0pt}}l@{\hspace{1em}}  r@{\hspace{1em}}r@{\hspace{1em}}r@{\hspace{1em}}r@{\hspace{1em}}r@{\hspace{1em}}}
\hline \\ [-1.2ex]
       &Transition &Freq.  &OCS&  \multicolumn{2}{l}{Time}  &   SNR   &  IFBW  &      Res.\tnote{a}\hspace{2mm} & $S$\tnote{b}\hspace{6mm} & OCS (cm$^3$)\hspace{1mm}  \\
       &  $J'\rightarrow J$   &(MHz)&(sccm)&  \multicolumn{2}{l}{(min)} &         &  (MHz) &     (kHz)   &    (MHz/min) & consumed\tnote{c} \hspace{1mm}  \\
\hline \\ [-1.5ex]

OC$^{36}$S:\\
bgc          &$2 \rightarrow 1$   &23198.73&0.25  &0&.33          &86. & 200.0   &   60.  &4,482,400&0.09\hspace*{4mm}\\
\textcolor{red}{\textit{bgc}}          &   &   &\textcolor{red}{\textit{0.01}} & &           &  &    &     &\textcolor{red}{\textit{7,200$^d$}} &\textcolor{red}{\textit{2.35$^d$}}\hspace*{4mm}\\
cavity       &$2 \rightarrow 1$   &23198.73&0.01  &0&.33          &36. & 0.5   &   2.  &  1,960&8.62\hspace*{4mm}\\
cavity       &$1 \rightarrow 0$   &11599.37&0.01  &0&.33          &31. & 0.5   &   2.  &  1,460&11.58\hspace*{4mm}\\
broadband    &$1 \rightarrow 0$   &11599.37&0.01  &28&. &13.      & 10000.0    &  50.   &60,360&0.28\hspace*{4mm}\\
\\
$^{18}$O$^{13}$C$^{34}$S:\\

bgc            &  $2 \rightarrow 1$ &22179.46&0.25 &50 &.  &5. & 200.0  &60. &100.0&   \\
            &    & &\textcolor{red}{\textit{0.02}} &  &  &  &    & &\textcolor{red}{\textit{0.6$^d$}} &\hspace{2mm}    \\
cavity         &  $2 \rightarrow 1$ &22179.46&0.02 &87 &. &6.  & 0.5  &2.   &0.2&   \\
cavity         &  $1 \rightarrow 0$ &11089.74&0.02 &87 &. &4. & 0.5  &2. &0.1&   \\
\hline 
\multicolumn{11}{l}{\small $^a$ Resolution.}\\
\multicolumn{11}{l}{\small $^b$ Spectral acquisition velocity $S$ = (SNR$^2$)(IFBW/Time).}\\
\multicolumn{11}{l}{\small $^c$ Quantity of OCS (cm$^3$) consumed in order to match the frequency coverage of the}\\
\multicolumn{11}{l}{\small \hspace{3mm}broadband experiment (10\,GHz, SNR 13, see text).}\\
\multicolumn{11}{l}{\small $^d$ Calculated assuming the buffer gas cell measurement was performed with the same flow rate } \\
\multicolumn{11}{l}{\hspace*{3mm}(either 0.1 or 0.2\,sccm) as the cavity and broadband meausurements.}\\
\end{tabular}
\end{threeparttable}
\end{table*}

Care must be taken when comparing $S$ values for the three different experiments, since this quantity depends on the flow rate of OCS. To more easily make comparisons, two rows are provided in Table \ref{OCStable} (in red) in which $S$ has been calculated  assuming an identical flow rate (either 0.01\,sccm or 0.02\,sccm of OCS) in all three experiments. In the case of the OC$^{36}$S measurement, decreasing the OCS flow rate by a factor of 25 in the buffer gas cell results in a factor of $(25)^2$ decrease in $S$ (4,482,400/7,200 = 625). Even at the same sample consumption rate, however, $S$ for the buffer gas cell is still five times greater than that of the cavity instrument, without accounting for the time required to repeatedly tune the cavity.

Another point of comparison is the total sample that would be consumed over a 10\,GHz wide spectral window, assuming the same SNR as the broadband instrument \cite{brown2008broadband}. 
For example, $S$ for the buffer gas cell is roughly 75 times greater than the broadband instrument, although sample is consumed 25 times faster. 
Since the broadband measurement required 28 minutes to achieve a SNR of 13 for the $J=1\rightarrow0$ transition of OC$^{36}$S, only 0.37\,min would be needed to achieve the same frequency coverage and SNR in the buffer gas cell. 
Under these conditions, the total consumption of OCS is 0.09\,cm$^3$, or about one-third the amount of OCS consumed in the 28\,min broadband experiment.  

This finding  illustrates a real strength of the sample delivery method in the buffer gas cell: the ability to use ``neat" or pure samples, i.e.~without the need for dilution.
Since high flow rates of buffer gas relative to the sample of interest are required to achieve low rotational temperatures in the pulsed jet experiments, it is often challenging to realize high sample consumption rates.
It should be emphasized that the  performance of the buffer gas cell can be achieved with modest electronics since high power TWTs, and high sampling rate digital oscilloscopes and AWGs are not required.

Our injection technique separates the cooling process from the chemistry, an operating regime which is quite unlike that of pulsed jet sources where these two processes are closely coupled (e.g., electrical discharge sources). 
For this reason, entirely new types of experiments may be possible, of which kinetic studies appear particularly promising. As recently demonstrated by Porterfield et. al. \cite{porterfield2018ozonolysis}, nascent product ratios can be determined within a few seconds following initiation of a gas phase chemical reaction. Beyond these types of studies, experiments to thermally arrest or ``titrate'' a chemical reaction as a function of time as a means to identify reactive intermediates may be feasible. Other studies that might be enabled by the present instrument include reactive collisions in the cold cell, such as barrierless radical chemistry, or the formation of van der Waals complexes.

\section{Conclusions}
\vspace{-4mm} Significant improvements have been made to a cryogenic buffer gas cooling cell with high resolution chirped pulse microwave electronics. 
By cryogenically cooling the first-stage amplifiers and protection switches, the system temperature (T$_{sys} \sim$ \,30\,-\,35\,K) has been considerably reduced relative to that of traditional room temperature microwave receivers ($T_{sys} >$ 300\,K).
With a moderate IFBW of 100\,-\,200\,MHz, a method for concatenating together individual frequency segments to produce a much wider bandwidth spectrum has been described. 
An extension of the operating range of the instrument, previously 12\,-\,18\,GHz and now 12\,-\,26\,GHz, has enabled a wider range of molecules to be detected.   
As demonstrated by detection of rare isotopic species of OCS in natural abundance, microwave spectra can be acquired with comparable detection sensitivity to traditional microwave cavity experiments, but several thousand times faster with moderate increases in sample consumption rate.
The performance of this instrument is also capable of exceeding that of wide-band CP spectrometers in terms of spectral acquisition velocity by nearly a factor of 100 with an increased rate of sample consumption.
As illustrated from a cursory analysis of lemon oil, this next generation instrument is a powerful and flexible analytical tool, capable of separating complex mixtures and detecting trace species with short acquisitions times and at high resolution.
\section{Supplementary Material}
\vspace{-4mm}
A detailed description of the microwave electronics following the labels presented in Figure 3 can be found in the Supplementary Information.

\section{Acknowledgements}
\vspace{-2mm}This work was supported by NSF Award DBI-1555781.  Additional support for McCarthy and Porterfield was provided by NSF Award CHE-1566266.\\ 

\section{Conflicts of Interest}
One author (DP) is an inventor on US patent 8,907,286, ``Gas phase cooling and mixture analysis'' The authors have no additional conflicts of interest to declare.
\vspace{-5mm}\section{Supplemental Information}
\vspace{-3mm}
The following is a list of components used in the 12-26\,GHz circuit of the cryogenic buffer gas cell described in this paper. Item parts are included with key attributes:

\begin{enumerate}
\item Homemade programmable TTL pulse generator. 
\item Arbitrary waveform generator; both RIGOL 4102 and 4202 have been used, with 100\,MHz and 200\,MHz bandwidth, respectively. 
\item Analog Devices GaAs Fundamental Mixer HMC773ALC3B, 6\,-\,26 GHz. Typical conversion loss of 9 dB, IF bandwidth DC to 8 GHz, RF to IF isolation of 20 dB typical. 
\item Marki Microwave Wideband Wilkinson Power Divider PD0426. In-phase power splitting 4\,-\,26.5\,GHz, typical output to output isolation 18 dB, 3 dB nominal power splitting, typical 0.8 dB insertion loss.
\item Mercury systems power amplifier L1826-32-T325, 18\,-\,26 GHz, 32 dBm saturation limit, small signal gain 32 dB min, noise figure 12 dB max.
\item Mercury systems power amplifier L1218-32-T325, 12\,-\,18 GHz, 32.5 dBm saturation limit, small signal gain 33 dB min, noise figure 8 dB max. 
\item A homemade complimentary level shifter with normal and inverse outputs, 0 and -5 V (could instead be achieved with analog switches). 
\item Standford Research Systems 4 channel delay generator, DG535, $<$\,50\,ps root mean square jitter, 5\,ps delay resolution.
\item Local oscillator Hewlett-Packard-8673D synthesized signal generator, 0.05-26 GHz.
\item Marki Microwave Wideband Wilkinson Power Divider PD0126. In-phase power splitting 1\,-\,26.5\,GHz, 20 dB typical output to output isolation, 3 dB nominal power splitting, typical 1 dB insertion loss. 
\item Analog Devices HMC383LC4 medium power amplifier 12\,-\,30 GHz, 15 dB gain, typical 8 dB noise figure. 
\item See (k). 
\item Pasternak (PE9854/SF-20) WR-62 Waveguide Standard Gain Horn Antenna 12.4\,-\,18 GHz, nominal 20 dB gain.  
\item Pasternak (PE9852/2F-20) WR-42 Waveguide Standard Gain Horn Antenna 18\,-\,26.5 GHz, nominal 20 dB gain. 
 \skipitems{1}
\item Low Noise Factory cryogenically cooled low noise amplifier LNF-LNC6\_20C, 6\,-\,20\,GHz, 32 dB gain, 0.070 dB typical noise figure, 4.7 K typical noise temperature. 
\item Low Noise Factory cryogenically cooled low noise amplifier LNF-LNC15\_29B, 15\,-\,29\,GHz, 32 dB gain, 0.125 dB typical noise figure, 8.5 K typical noise temperature.
\item Narda-MITEQ Amplifier JS4-00102600-30-10P, 0.1\,-\,26\,GHz, 28 dB gain, typical noise figure 3 dB. 
\item Analog Devices low noise amplifier HMC962LC4, 7.5\,-\,26.5\,GHz, 13 dB gain, low noise figure 2.5 dB. 
\item See (k). 
\item See (c). 
\skipitems{1}
\item Stanford Research Systems SR445A preamplifier, DC-350 MHz, four wide-bandwidth, DC-coupled amplifiers with a gain of 5 each. When four channels used as cascade, total voltage gain is 625. 
\item Keysight U1084A data acquisition card, 8-bit PCIe high-speed digitizer with on-board signal processing, 128 MB processing memory, real-time sampling and averaging, 2 GS/s with 2 channels. 
\skipitems{1}
\item Analog Devices HMC547ALC3 GaAs single-pole, single-throw non-reflective switch, DC\,-\,28 GHz. 

\end{enumerate}

\end{document}